\documentclass[9pt,twocolumn,twoside]{opticajnl}
\journal{opticajournal} 

\setboolean{shortarticle}{true}

\usepackage{comment}
\usepackage{lineno}

\usepackage{bm}
\usepackage{graphicx}
\title{{Planar near-field measurements of specular and diffuse reflection of millimeter-wave absorbers}}

\author[1,2,*]{Fumiya Miura}
\author[2]{Hayato Takakura}
\author[2]{Yutaro Sekimoto}
\author[2]{Junji Inatani}
\author[2]{Frederick Matsuda}
\author[2]{Shugo Oguri}
\author[1]{Shogo Nakamura}

\affil[1]{Department of Physics and Engineering, Graduate School of Science and Engineering, Yokohama National University, 79-1, Tokiwadai, Hodogaya-ku, Yokohama, Kanagawa 240-8501, Japan}
\affil[2]{Institute of Space and Astronautical Science (ISAS), Japan Aerospace Exploration Agency (JAXA), 3-1-1 Yoshinodai, Chuo-ku, Sagamihara, Kanagawa 252-5210, Japan}

\affil[*]{miura-fumiya@ac.jaxa.jp}

\begin{abstract}
Mitigating the far sidelobes of a wide field-of-view telescope is one of the critical issues for polarization observation of the cosmic microwave background. 
Since even small reflections of stray light at the millimeter-wave absorbers inside the telescope may create nonnegligible far sidelobes, we have developed a method to measure the reflectance of millimeter-wave absorbers, including diffuse reflections.
By applying the planar near-field measurement method to the absorbers, we have enabled two-dimensional diffuse-reflection measurements, in addition to characterizing specular reflection. 
We have measured the reflectance of five samples (TK RAM Large and Small Tiles and Eccosorb AN-72, HR-10, and LS-22) at two angles of incidence in the frequency range from 70 GHz to 110 GHz. 
Compared with conventional horn-to-horn measurements, we obtained a consistent specular reflectance with a higher precision, less affected by standing waves.
We have demonstrated that the angular response and diffuse-to-specular reflectance ratio differ among various materials. The measurements also imply that some absorbers may affect the polarization direction when reflecting the incident waves.
\end{abstract}

\setboolean{displaycopyright}{false} 

\begin{document}

\maketitle

\section{Introduction}
\par
The primordial gravitational waves due to inflation are believed to have left a characteristic polarization pattern called $B$ modes in the cosmic microwave background (CMB)\cite{Kamionkowski2016bmode}. 
The $B$-mode polarization is expected to peak at large angular scales of several degrees or larger, and the stray light contamination due to emission from the Galactic plane entering the telescope is expected to be one of the largest systematic errors\cite{planckLFI2001, E.Hivon2017, H.Tran2010}. 
Stray light can enter the telescope through the far sidelobes. Next-generation CMB telescopes such as \textit{LiteBIRD} need to know sidelobe to an accuracy of -56 dB\cite{PTEP2022}. 
\par
To achieve such a requirement, it is essential to characterize reflections from millimeter-wave absorbers\cite{Takakura2022}, for example, at the baffling walls, in addition to minimizing stray light in the optical design optimization\cite{Sekimoto2020}.
In recent years, novel millimeter-wave absorbers have been reported, such as metamaterial absorbers\cite{Pisano:23, Xu:21}, 3D printed absorbers\cite{Petroff2019, Adachi2020, Otsuka:21}, and new carbon-based absorbers\cite{Inoue:23, Yanagi_2021}. 
\par
The reflectance properties of millimeter-wave absorbers in free space have been made using two horns and focusing optics\cite{Xu:21} in such a way that the attenuation of the optical coupling between the horns is measured (\textit{horn-to-horn} method).
Since stray light enters the absorber from various angles inside the telescope, the specular reflectance is measured at multiple angles of incidence\cite{Blanco1985, Norouzian2016}.
Also, by rotating one of the horn antennas around the absorbers, one-dimensional distributions of diffuse reflection have been measured\cite{saily2003, Saily2004, Xu:21}. However, if the scattered waves from the absorber are only partially coupled to the receiving horn, the measured reflectance is underestimated.
\par
In this paper, we report a new method to measure the reflectance of millimeter-wave absorbers. Applying the planar near-field antenna measurement method\cite{Yaghjian1986, Parini2008, Raisanen2018}, we measure the amplitude and phase of the reflected waves from absorbers on a plane. By applying plane wave expansion to the measurements, we characterize the two-dimensional angular responses of diffuse reflection, from which the specular reflectance is also evaluated.
\par
We characterized the specular and diffuse reflectance for five absorbers (AN-72, HR-10, LS-22, TK RAM Large, and TK RAM Small) in the 70 -- 110 GHz range. 
Since the reflectance characteristics of the absorber depend on the polarization direction\cite{orfanidis2016, saily2003}, we performed measurements for both incident polarizations (S-pol and P-pol). We set the polarization direction of the probe not only parallel but also perpendicular to the polarization direction of the feed horn.
\par
The structure of this paper is as follows. 
Section~\ref{measurementmethod} describes the principles of the new measurement method proposed in this paper. 
Section~\ref{experimentalverification} describes the measurement setup constructed to demonstrate the measurement method. 
Section~\ref{results} reports the measurement results of millimeter-wave absorbers and discusses the specular and diffuse reflectance. 
Section~\ref{conclusion} states the conclusion.

\section{Reflection measurement method}
\label{measurementmethod}
\subsection{Horn-to-horn method}
\label{horntohorn}
\begin{figure*}
\centering\includegraphics[width=14cm]{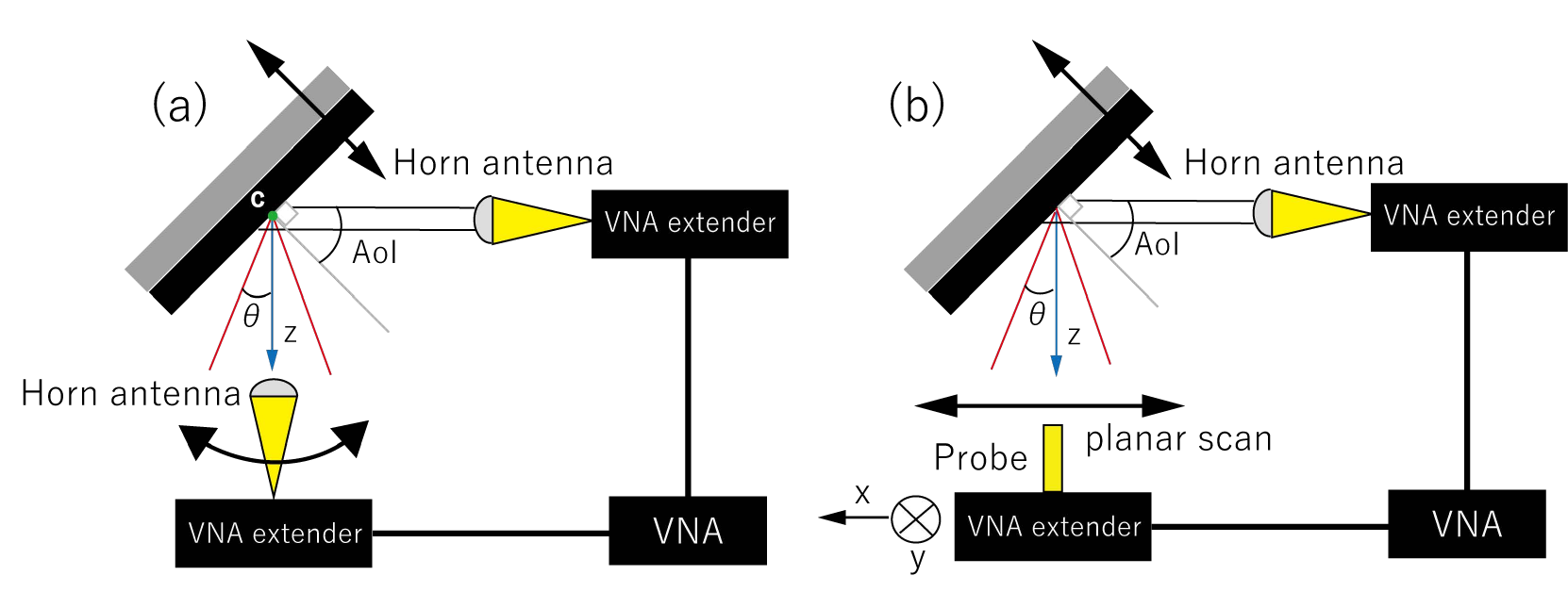}
\caption{Schematic of reflection measurements of millimeter-wave absorbers. 
(a)~Conventional\cite{saily2003, Norouzian2016, Xu:21} measurement method (horn-to-horn method). A horn antenna illuminates the absorber, and the optical coupling between the horns is measured. One-dimensional diffuse reflection is also measured by rotating the receiving horn concerning the reflection point. 
The reflection point and the center of rotation must coincide. 
Point c indicates the center of rotation.
The blue lines show specular reflection, and the red lines show diffuse reflection.
(b)~Planar near-field measurement method. We illuminate the test absorber and measure the amplitude and phase of the reflected waves by moving the receiver with the probe on a two-dimensional plane. The measured amplitude and phase distributions can be separated into angular plane wave components. $z$-axis represents the propagation direction of the specular reflection. The angle from the $z$-axis was defined as $\theta$.}
\label{method}
\end{figure*}

Conventional millimeter-wave absorber reflection measurements have been made using two horn antennas and measuring the attenuation of the optical coupling between the horn antennas (\textit{horn-to-horn} method; Fig.~\ref{method}(a))\cite{Norouzian2016, Xu:21}. Since the optical coupling should be as high as possible for accurate measurements, the horn antennas are often used with focusing optics, such as mirrors\cite{Xu:21} and lenses.
This method has been extended to one-dimensional diffuse reflection characterization by changing the angle of the receiving horn ($\theta$ in Fig.~\ref{method}(a))\cite{saily2003, Saily2004}.  
However, if the scattered waves from the absorber are only partially coupled to the receiving horn, the measured reflectance is underestimated. It has been reported that the measured reflectance depends on the distance between the absorber and the receiving horn\cite{saily2003}.

\subsection{Planar near-field method}
To characterize two-dimensional specular and diffuse reflections of millimeter-wave absorbers, we propose to apply the planar near-field antenna measurement method to the reflection measurement of millimeter-wave absorbers.
The planar near-field measurements \cite{Yaghjian1986, Parini2008, Raisanen2018} derive the angular radiation pattern $I(\theta,\phi)$ of an antenna by the Fourier transformation of the complex electric field at the aperture of the antenna, based on the Fraunhofer diffraction \cite{BornWolf1999} or van Cittert-Zernike theorem \cite{Thompson2017}.
The complex electric field at the aperture is measured using a vector network analyzer by scanning a probe in a specific scan plane.
To measure the field with wide angles, a small-aperture probe is used, such as an open-ended waveguide.
The measured amplitude and phase distributions are decomposed into plane-wave components with different angular directions $\varepsilon$ by two-dimensional Fourier transform as,  
\begin{equation}
    I(\theta,\phi) = |\varepsilon(k_l,k_m)|^2; 
    \quad
    \varepsilon (k_l,k_m) \propto \iint E(x,y)e^{-i(k_lx+k_my)}dxdy.
    \label{farfield}
\end{equation}
Here, $E (x,y) = |E (x,y)| \exp(i\Phi (x,y)) $ is the complex electric field measured near the test antenna aperture; $(k_l,k_m)$ denote the spatial frequencies in the $(x,y)$ plane, which relate the angle $(\theta,\phi)$ as  $k_l=k\sin{\theta}\cos{\phi}$ and $k_m=k\sin{\theta}\sin{\phi}$ with $k$ being the wave number; $\theta$ and $\phi$ correspond to the angle from the $z$-axis (perpendicular to the $x$--$y$ plane) and the azimuth angle from the $x$-axis, respectively.
\par
By applying the planar near-field measurement method to the reflection measurement of millimeter-wave absorbers, the angular response of the reflected waves from the absorbers, including diffuse reflection, can be acquired. 
In this method, we illuminate the test absorber with a horn antenna and measure the waves reflected from the test absorber by moving the receiver with the probe on a two-dimensional plane ($x, y$) (Fig.~\ref{method}(b)). The measured amplitude and phase distributions $E (x,y)$ are decomposed into the angular response $I(\theta,\phi)$ using Eq.~(\ref{farfield}). Here, we defined the propagation direction of the specular reflected wave as the $z$-axis and set the scan plane ($x, y$) perpendicular to the $z$-axis. 
The $y$-axis is set in the S-polarization direction. Note that P-polarized waves oscillate with the electric field parallel to the plane of incidence, while S-polarized waves oscillate perpendicular to the plane of incidence.
\par
The planar near-field measurement has three advantages.
First, by measuring the reflected wave of the absorber, we can evaluate the reflectance characteristics in more detail. Compared with the conventional horn-to-horn method, the information obtained from the two-dimensional distribution of reflected waves, including diffuse reflections, is much greater.
Second, since there is no need to know the reflection point of the absorber, it is less susceptible to alignment adjustments in optical measurements.
Third, compared with horn-to-horn measurements, near-field measurements are less affected by standing waves because of the small aperture probe with a small optical coupling.
\par
We evaluate the specular and total reflectance of the absorbers by integrating the measured two-dimensional angular responses into different integration ranges. 
The former is obtained by the integration of the reflected waves in the half width at half maximum (HWHM) of the beams of the horn antenna,
\begin{equation}
    R^{\text{s}} = \frac{\int^{\theta=\text{HWHM}}_{\theta=0^\circ}I_{\text{absorber}}(\theta,\phi)d\Omega}{\int^{\theta=\text{HWHM}}_{\theta=0^\circ}I_{\text{PEC}}(\theta,\phi)d\Omega},
    \label{specular}
\end{equation}
which is considered to be similar to the conventional horn-to-horn measurement.
The latter is obtained by the integration of the entire two-dimensional angular reflectance ($0^\circ \leq \theta \leq \theta_{\text{max}}$), 
\begin{equation}
    R^{\text{t}} = \frac{\int^{\theta=\theta_{\text{max}}}_{\theta=0^\circ}I_{\text{absorber}}(\theta,\phi)d\Omega}{\int^{\theta=\theta_{\text{max}}}_{\theta=0^\circ}I_{\text{PEC}}(\theta,\phi)d\Omega}.
    \label{diffuse}
\end{equation}
Here, $I_{\text{PEC}}(\theta,\phi)$ and $I_{\text{absorber}}(\theta,\phi)$ are the angular responses calculated from Eq.~(\ref{farfield}) for the perfect electric conductor (PEC) and test absorber, respectively. For the case of our experimental verification described in Sections \ref{reference} and \ref{angularresponse}, the HWHM and $\theta_{\text{max}}$ are $4.3^\circ$ and $30^\circ$, respectively.

\section{Experimental verification}
\label{experimentalverification}
\subsection{Experimental setup}
\label{experimentalsetup}
\begin{figure}
\centering\includegraphics[width=7cm]{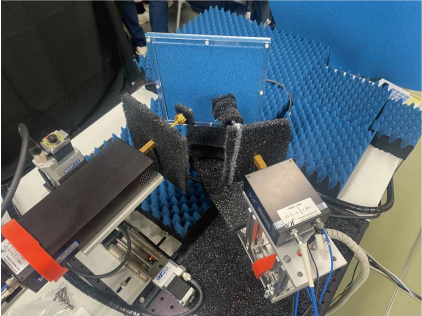}
\caption{A photograph of the measurement setup corresponding to Fig.~\ref{method}(b).}\label{setup}
\end{figure}

Figure~\ref{setup} shows the measurement setup developed in this study. We measured the amplitude and phase of the electric field reflected by the test absorbers using a vector network analyzer (VNA: Keysight Technologies N5222B) and frequency extenders (Virginia Diodes WR10VNAX)\cite{Takakura2019}. Based on the reciprocity theorem, the optical coupling between the two horns is equivalently measured when the VNA transmitter and receiver are reversed\cite{Viktar2020}. For the convenience of the setup, we connected a probe and a lens horn to the transmitter and receiver, respectively, in this experiment. 
\par
The measurements were performed with a 1 GHz interval over a frequency range of 70 GHz -- 110 GHz. Signals from 11.6 to 18.3 GHz were output from the VNA and up-converted to six times higher frequencies. Then, the 70 -- 110 GHz signal was output from the probe. We used an open-ended waveguide as the probe\cite{naruse2009}. We measured the reflected waves from the millimeter-wave absorber by scanning the probe in 1.5 mm increments in a 75 mm $\times$ 75 mm square area. The distance between the probe and test absorbers was set to 60 mm, considering the mechanical clearance. This configuration measures the specular and diffuse reflection in $0^\circ \leq \theta \leq 30^\circ$. The reflected waves were received by a lens horn, which consists of a conical horn (aperture diameter 23.0 mm, opening angle 10.0$^\circ$) with a polymethyl pentene (TPX) lens (diameter 25.4 mm, focal length 50 mm) placed in the aperture. 
The distance between the lens horn and test absorbers was 50 mm. Based on a Gaussian-beam approximation\cite{goldsmith}, the beam radius at the absorber is 10.7 mm.
\par
The millimeter-wave absorber under test was placed on two rotating and one-axis stages.
Since the movable range of the measurement stage was finite, in addition to the angle of incidence of the lens horn (AoI in Fig.~\ref{method}(b)), the angle of the absorber relative to the measurement stage was also rotated so that the measurement plane ($x$--$y$ plane) and the signal optical path ($z$-axis) becomes perpendicular. In addition, using a one-axis stage, we aligned the surface of the absorbers at a common plane for all the measurements.
Considering the situation in telescopes, we measured the absorbers fixed on a 5 mm thick aluminum plate. The absorbers and the aluminum plate were 200 mm $\times$ 200 mm in size.
An acrylic frame held the absorber in place, leaving no space between it and the aluminum plate.
\par
Table~\ref{tbl:measured-parameter} summarizes the measurement parameters for this experiment. We measured the reflectance for two polarizations; the polarization direction of the lens horn was set parallel to the $y$ and $x$-axes in Fig.~\ref{method} (S- and P-pol, respectively). We checked the mechanical accuracy of the coordinates using a level gauge.
In addition, the polarization direction of the probe was set not only parallel but also perpendicular to that of the lens horn.

\begin{table*}
    \begin{center}
        \caption{Near field measurement parameters. 
        The polarization direction of the S- and P-polarized lens horn are parallel to the $y$ and $x$-axes in Fig.~\ref{method}, respectively.
        The ``sp'' polarization represents the S-polarized lens horn and the P-polarized probe, and vice versa for ``ps''.}
        \label{tbl:measured-parameter}
        \begin{tabular}{lc}
        \hline 
        Frequency range (step) & 70 - 110 GHz (1 GHz)\\ 
        Scan range (step) & 75 mm $\times$ 75 mm (1.5 mm)\\
        Angle of incidence (AoI)  &  30 deg, 45 deg \\  
        Co-polarization of the lens horn & S-pol, P-pol \\
        Probe polarization & ss, pp (parallel to the lens horn)\\& sp, ps (perpendicular to the lens horn) \\
        \hline
        \end{tabular}
    \end{center}
\end{table*}

\subsection{Millimeter-wave absorbers}
\begin{table*}
\caption{Millimeter-wave absorber samples. An aluminum plate is placed on the back side of the absorbers during measurement.}
\begin{center}
\label{tbl:absorbers}
\begin{tabular}{lcccccc}
\hline
Samples & Material & Form  & Structure & Thickness & Mass & Ref.\\
&&&&[mm]&[kg/m$^2$]&
\\
\hline
AN-72 &polyurethane & sheet & multi-layer & \multicolumn{1}{r}{6} & \multicolumn{1}{r}{1.0}& \cite{an}\\
HR-10  &polyurethane & sheet & single-layer & \multicolumn{1}{r}{10} & \multicolumn{1}{r}{0.6}&\cite{hr}\\
LS-22 &polyurethane & sheet & single-layer & \multicolumn{1}{r}{9.8} & \multicolumn{1}{r}{0.5} & \cite{ls}\\
TK-L  &polypropylene & tile & pyramid  & \multicolumn{1}{r}{14.9} & \multicolumn{1}{r}{10.6}& \cite{thomaskeating}\\
TK-S &polypropylene & tile & pyramid  & \multicolumn{1}{r}{7.6} & \multicolumn{1}{r}{5.9}& \cite{thomaskeating}\\
\hline
\end{tabular}
\end{center}
\end{table*}

Table \ref{tbl:absorbers} summarizes the test millimeter-wave absorbers measured in this experiment.
Eccosorb AN-72\cite{an}, HR-10\cite{hr}, and LS-22\cite{ls} are commercially available millimeter-wave absorbers by Laird. They are made of polyurethane-based foams with different structures, which are expected to have different diffuse reflection characteristics: LS-22 has a single-layer structure; AN-72 has a multi-layer structure; HR-10 has a single-layer structure and contains carbon powder. LS-22 is available in several thicknesses, and we measured a 3/8-inch one. Eccosorb AN-72 and Eccosorb HR-10 are used in CMB ground-based experiments\cite{Sobrin_2022, 2022QUBIC, Takahashi2010}. 
TK RAM is a radar absorber developed by Thomas Keating for the terahertz and millimeter-wave bands, and there are two types of TK RAM\cite{thomaskeating}. 
TK RAM Small tile (abbreviated as TK-S) is a small tile of 25 mm $\times$ 25 mm and is recommended for 200 -- 600 GHz.
TK RAM Large tile (abbreviated as TK-L) is a large tile of 100 mm $\times$ 100 mm and recommended for 50 -- 200 GHz.
TK-L and TK-S have different-sized surface structures with a pyramidal shape, their heights being 5.0 mm and 2.5 mm, respectively.

\subsection{Reference measurement with an aluminum plate}
\label{reference}
For the intensity calibration, we measured the angular response of an aluminum plate without the absorber. Assuming that aluminum is a perfect electric conductor\cite{Serov2016}, the angular response of the aluminum plate $I_{\text{Al}} (\theta, \phi)$ is equivalent to that of the perfect electric conductor ($I_{\text{PEC}} (\theta, \phi)$ in Eqs.~(\ref{specular}) and (\ref{diffuse})). Hereafter, we use $I_{\text{Al}}$ instead of $I_{\text{PEC}}$.
\par
\begin{figure}
\centering\includegraphics[width=8cm]{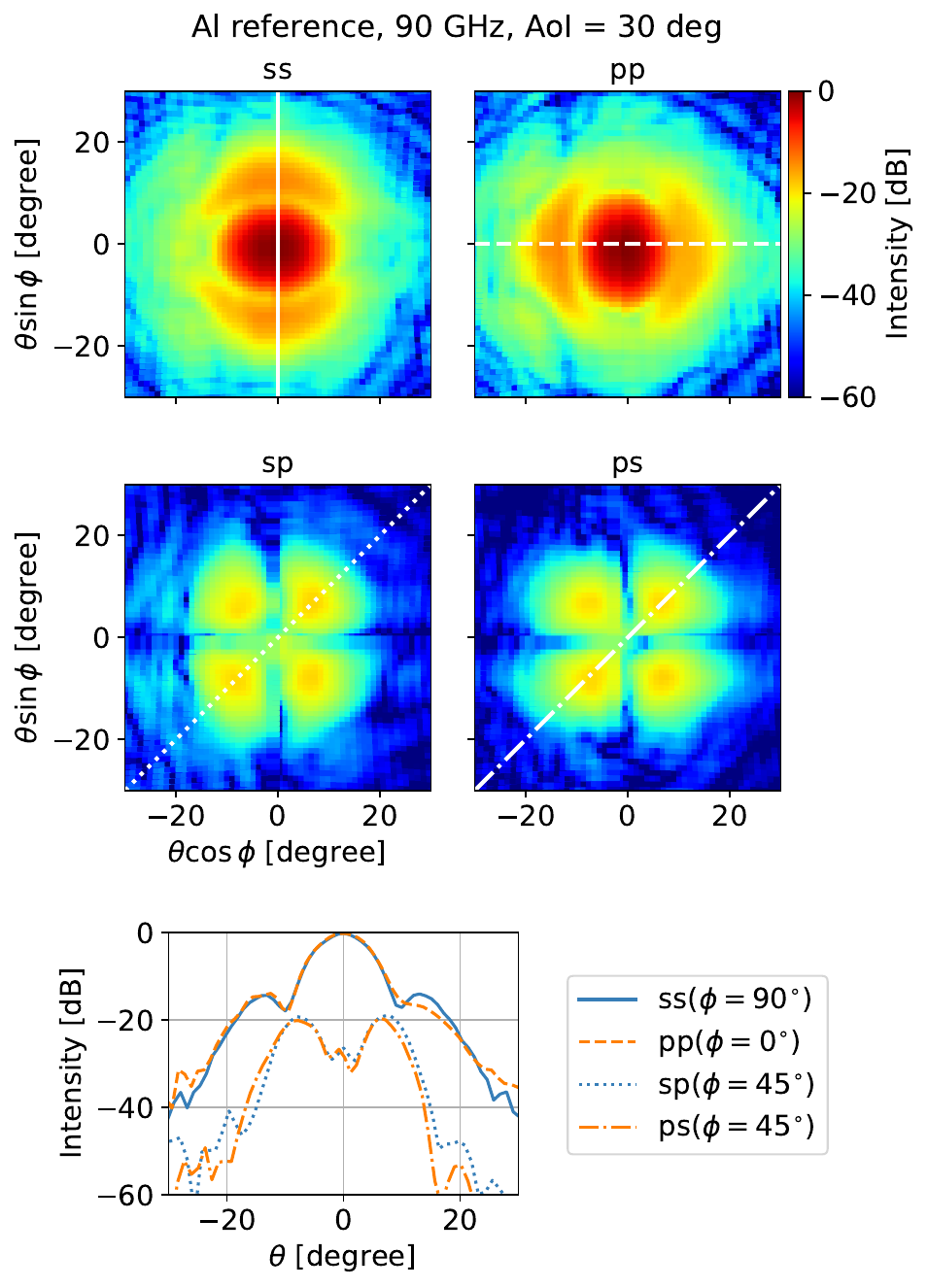}
\caption{Reflection measurements of the reference aluminum plate at 90 GHz for AoI = 30$^{\circ}$. The upper panels show the angular response of the electric field parallel and perpendicular to the co-polarization of the S- and P-polarized lens horn. Then, $\theta=0^\circ$ corresponds to the direction of specular reflection. These patterns are consistent with the antenna patterns of the lens horn used in this experiment. The lower panel shows $\phi$-cut profiles for parallel field ($\phi=90^\circ$ for ss $\phi=0^\circ$ for pp), and perpendicular field ($\phi=45^\circ$). The white lines in the two-dimensional patterns indicate the cross-sectional positions whose $\phi$ cuts are shown in the bottom panel. The angular response $I_{\text{Al}}(\theta,\phi)$ are calculated using Eq.~(\ref{farfield}), and $I_{\text{Al}}(\theta,\phi)$ is normalized by their peak value.}
\label{cal_Al}
\end{figure}

Figure~\ref{cal_Al} shows the reflection measurements of the reference aluminum plate at 90 GHz, measured with the electric field parallel and perpendicular to the co-polarization of the S and P-polarized lens horn.
The angular response $I_{\text{Al}}(\theta,\phi)$ are calculated using Eq.~(\ref{farfield}) and are normalized by their peak value. Here, we confirmed that the angular response of the reference aluminum measurement peaks at $\theta = 0^\circ$.
\par
These patterns are consistent with the lens horn pattern. The lens horn has larger sidelobes in the electric-field direction of the co-polarization pattern and quadrupoles in the cross-polarization pattern. Therefore, as expected, the ``ss'' and ``pp'' patterns have sidelobes in each polarization direction, and the ``sp'' and ``ps'' patterns show four lobes.
The HWHM of the lens horn was measured to be 4.3 $^\circ$ at 90 GHz, consistent with the prediction by a Gaussian-beam approximation\cite{goldsmith}.
The cross-polarization level of the lens horn was measured to be $-20$ dB at 90 GHz.
Since the patterns measured with both polarizations are consistent at the $-40$ dB level over a range of $0^\circ \leq \theta \leq 30^\circ$, we confirmed that this setup can measure two polarizations.

\section{Results and discussion}
\label{results}
\subsection{Planar near-field reflection measurements}
\begin{figure*}
\centering\includegraphics[width=\linewidth]{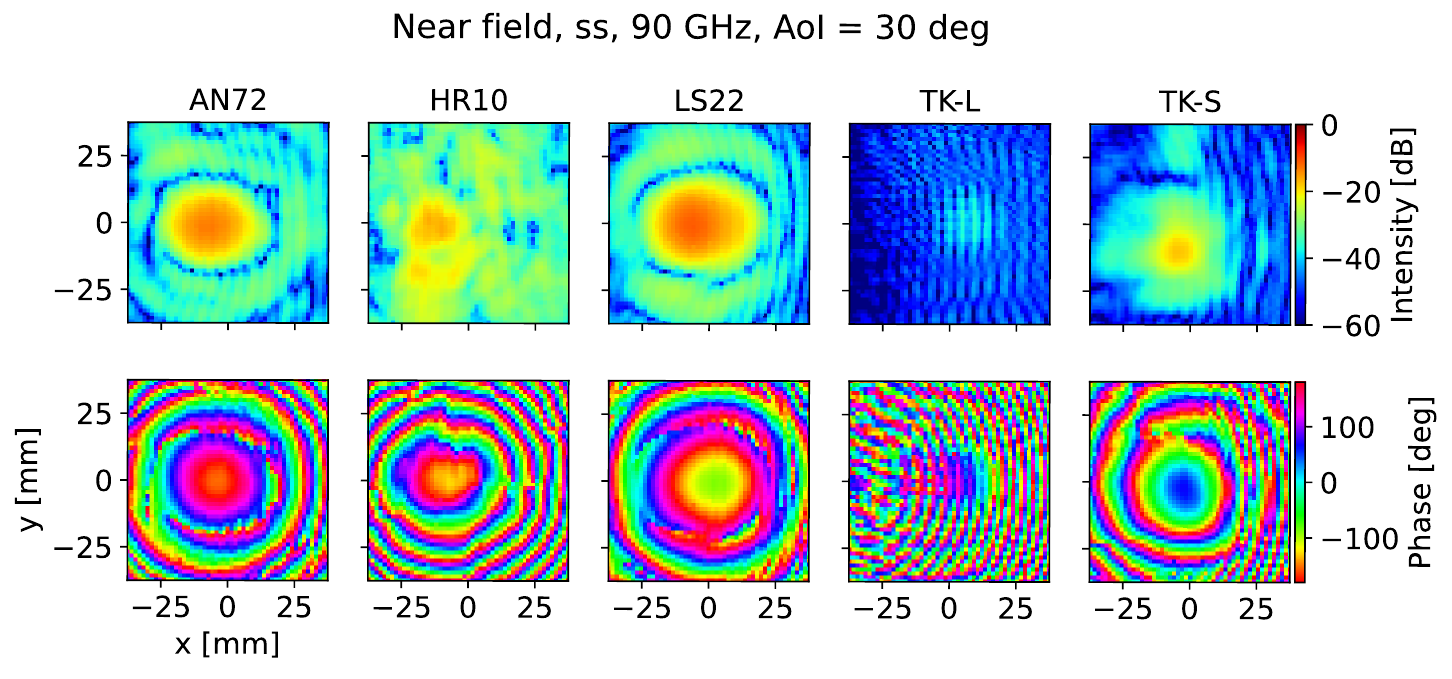}
\caption{Amplitude and phase distributions of the reflected waves from the millimeter wave absorbers $E(x,y)$ with ss polarization at 90 GHz for AoI = 30$^{\circ}$.
The upper panel shows amplitude, and the lower panel shows phase. Amplitudes are normalized by the peak amplitude of the reference aluminum plate.}
\label{nearfield}
\end{figure*}

We have measured the reflectance of the millimeter-wave absorbers using the planar near-field measurement with the setup described in Sec.~\ref{experimentalsetup}. 
Figure~\ref{nearfield} shows the amplitude and phase distributions of the measured millimeter-wave absorber reflections $E (x, y)$ with ss polarization (S-polarized lens horn; probe polarization parallel to the lens horn) at 90 GHz for AoI = 30$^{\circ}$.
The amplitude distributions are normalized by the peak value of the amplitude distribution of the reference aluminum plate for each polarization. 
Compared with AN-72, HR-10 sample has a more spread pattern.

\subsection{Angular responses and reflectance of the absorbers}
\label{angularresponse}
\begin{figure*}
\centering\includegraphics[width=\linewidth]{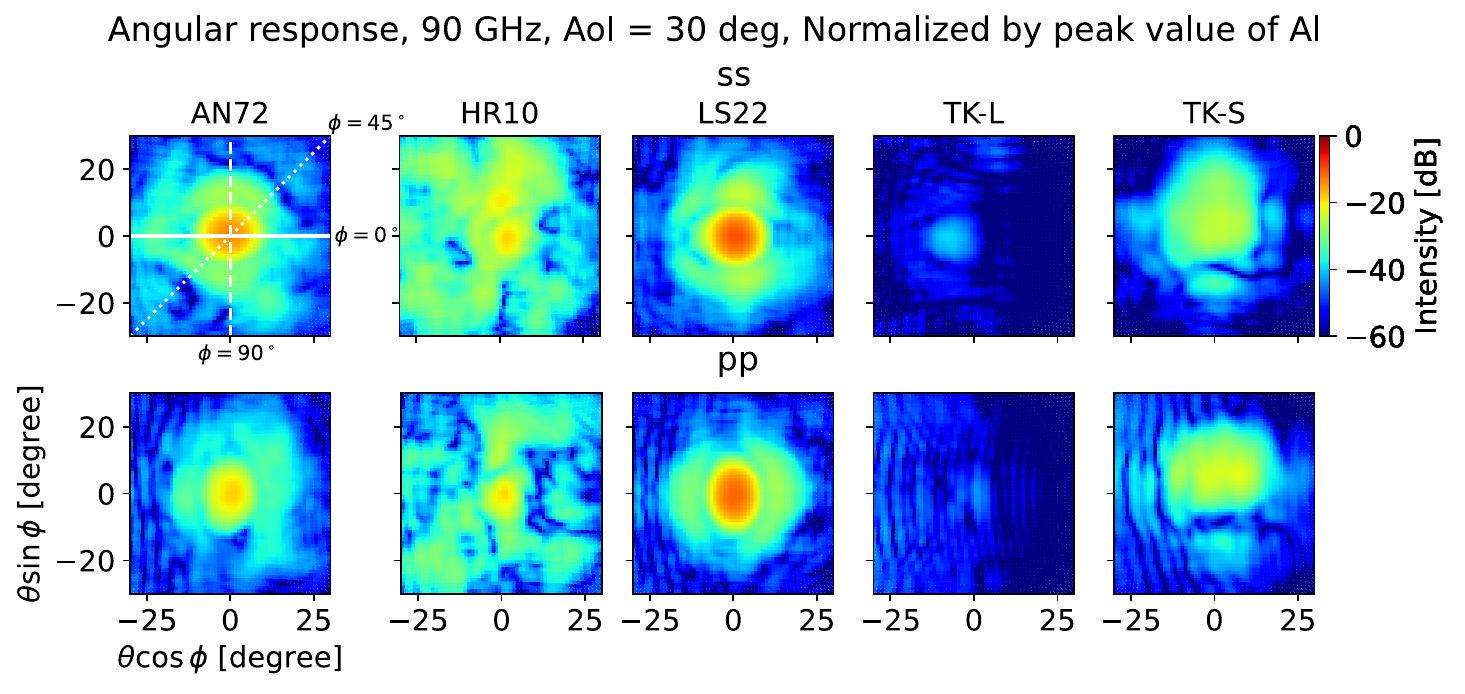}
\caption{Two-dimensional reflection patterns of reflectance of millimeter-wave absorbers $I_{\text{absorber}}(\theta, \phi)$ at 90 GHz for AoI = 30$^{\circ}$.
This pattern was calculated using the amplitude and phase of the reflected wave for each absorber by Eq.~(\ref{farfield}). 
The upper panel shows the measurements with ss polarization, and the lower panel shows the measurements with pp polarization. The white lines in the left-top panel indicate the cross-sectional positions whose $\phi$ cuts are shown in Fig.~\ref{phicut_30_co}. The white solid, dashed, and dotted lines correspond to $ \phi=0^\circ, 90^\circ$ and $45^\circ$, respectively.}
\label{farfield_30}
\end{figure*}
\begin{figure*}
\centering\includegraphics[width=\linewidth]{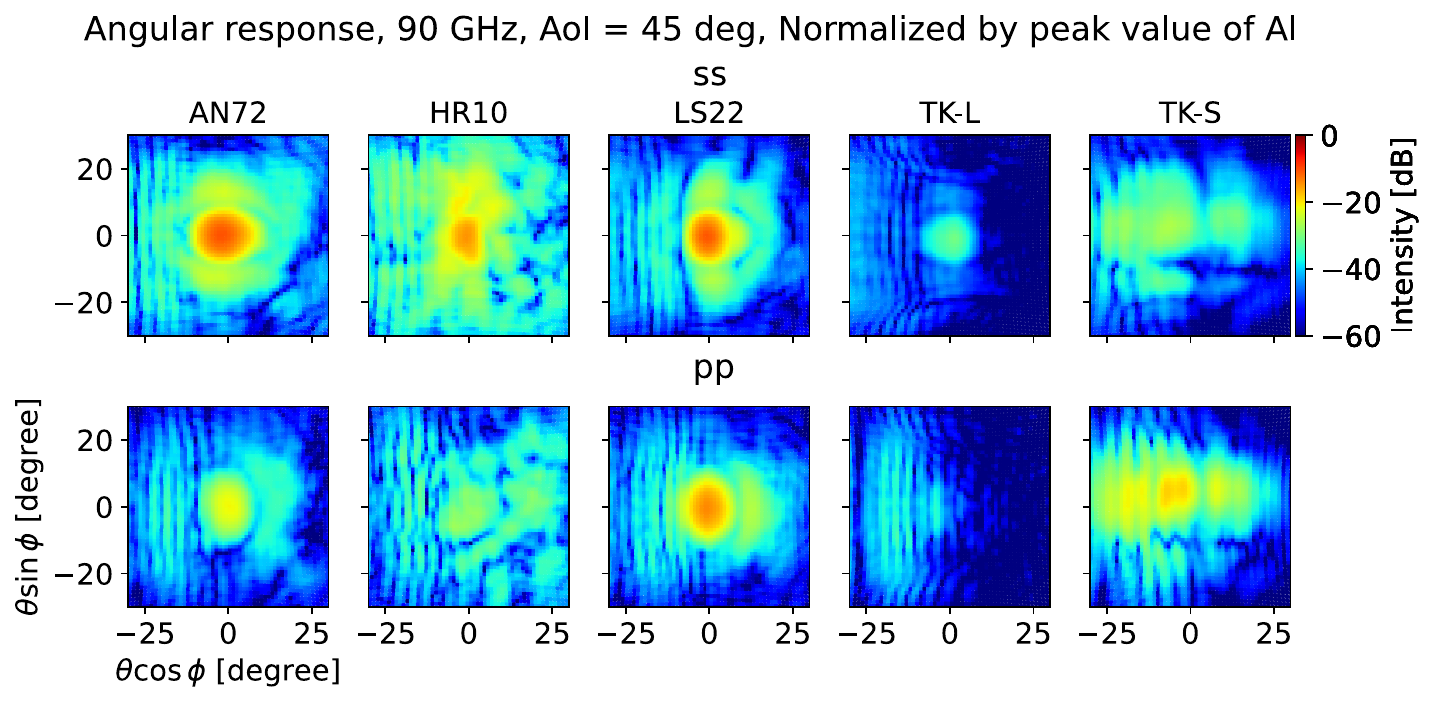}
\caption{Same as Fig.~\ref{farfield_30}, but shows the measurements for AoI = 45$^\circ$.}
\label{farfield_45}
\end{figure*}

Figures~\ref{farfield_30} and \ref{farfield_45} show the angular responses of the reflected waves from the absorbers $I_{\text{absorber}}(\theta, \phi)$ at 90 GHz for AoI = 30$^{\circ}$ and $45^{\circ}$, respectively.
The results are normalized by the peak value of the reference aluminum plate for each polarization.
AN-72 has an almost isotropic attenuation from the reference aluminum plate pattern, while HR-10 has a spread pattern.
LS-22 has similar patterns to AN-72, but the difference in S-pol and P-pol is smaller than that of AN-72. 
TK-S also shows a pronounced diffuse reflection pattern, indicating an upward tilt. 
This may be caused by the surface structure and errors in attaching TK-S to the reference aluminum plate.
\par
\begin{figure}
\centering\includegraphics[width=\linewidth]{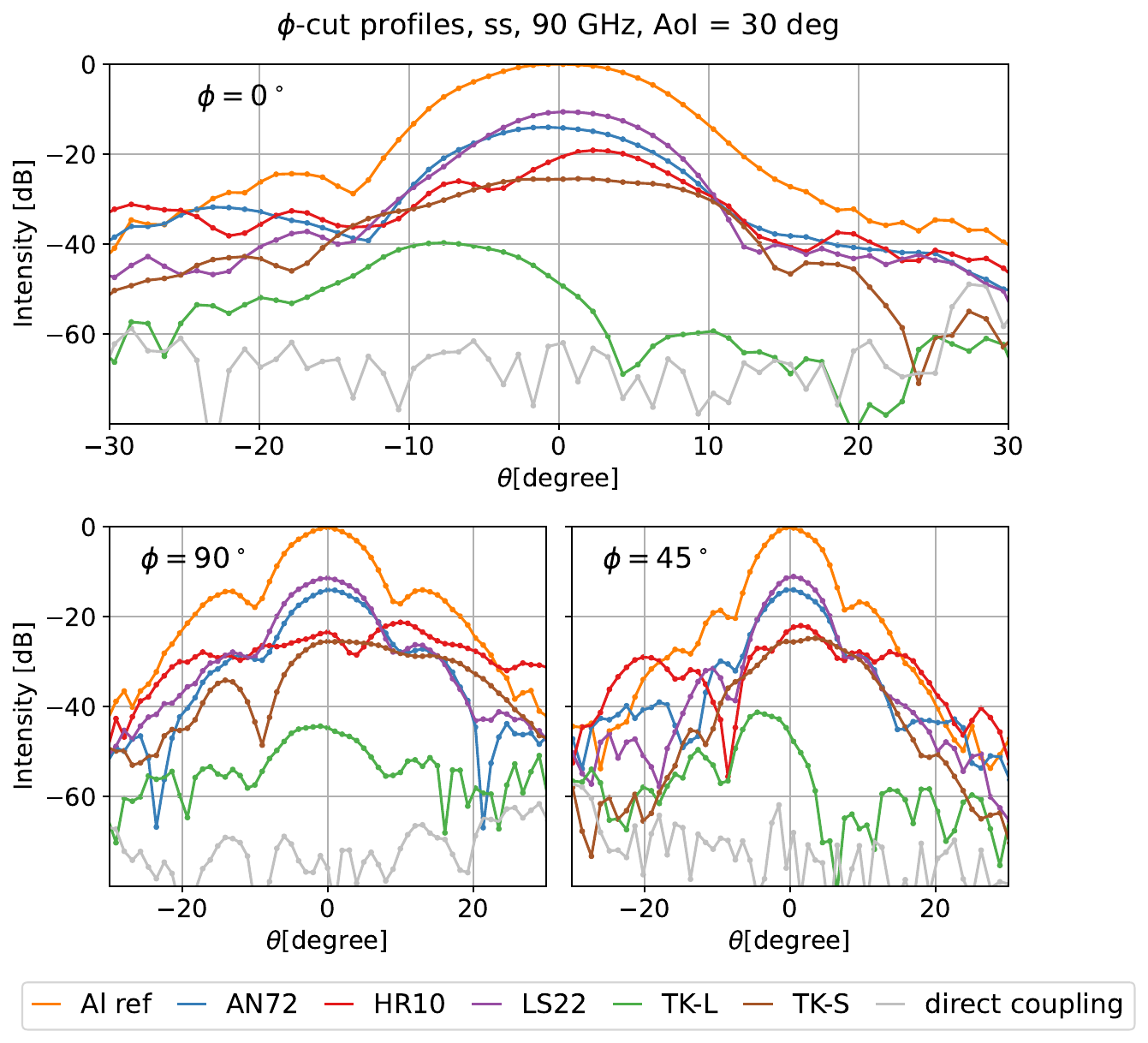}
\caption{Cross-sectional profiles of the measurements for AoI $= 30^\circ$ (Fig.~\ref{farfield_30}) in the $\phi=0^\circ, 90^\circ$ and $45^\circ$ directions.
The gray line shows the direct coupling discussed in Sec.~\ref{directpath}.}
\label{phicut_30_co}
\end{figure}

Figure~\ref{phicut_30_co} is the cross-sectional profiles of the measurements for AoI $= 30^\circ$ (Fig.~\ref{farfield_30}) in the $\phi=0^\circ, 90^\circ$ and $45^\circ$ directions.
While AN-72 and LS-22 have a constant attenuation from the reference aluminum plate pattern, the intensity of the reflected waves from HR-10 ranges between -40 and -20 dB regardless of the angle $\theta$.
Notably, in the angular range of 20$^\circ \leq \theta \leq$ 30$^\circ$, the level of the reflected waves from HR-10 is higher than that of the reference aluminum plate.

\begin{figure*}
\centering\includegraphics[width=15cm]{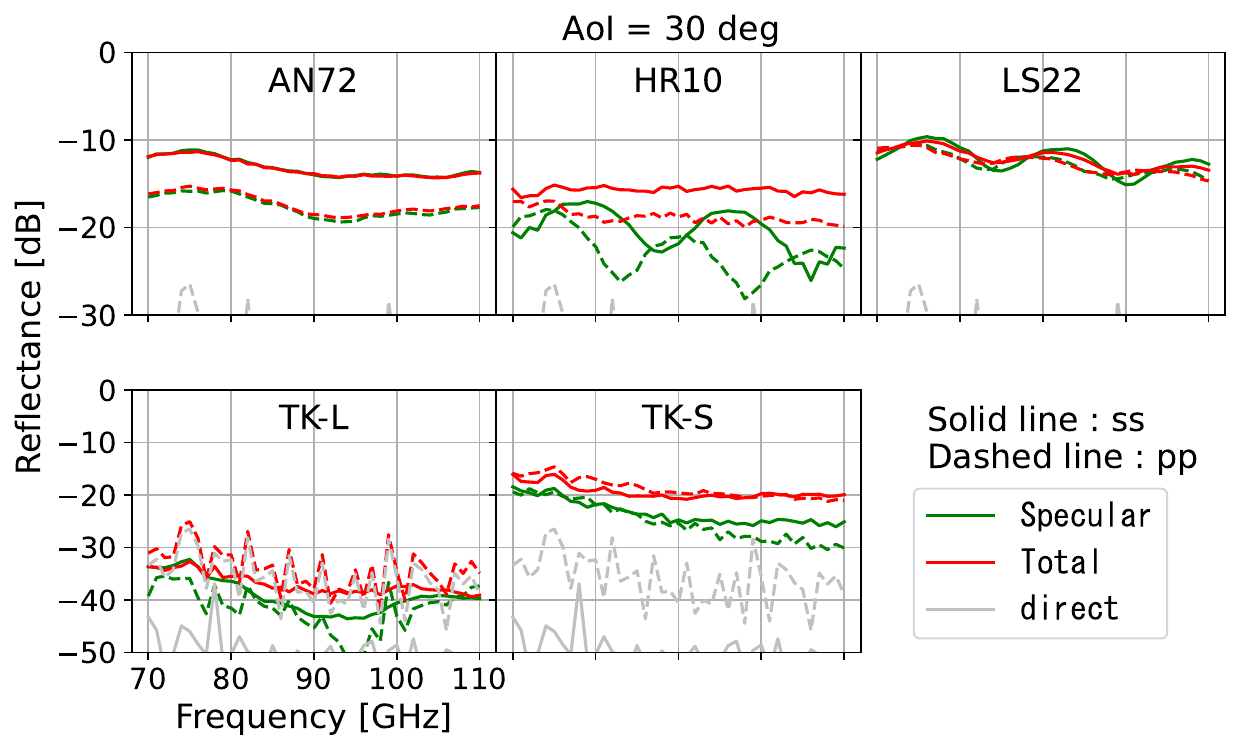}
\caption{Specular and total reflectance of millimeter-wave absorbers, $R^\text{s}$ and $R^\text{t}$, for AoI = 30$^\circ$. 
The solid lines represent the measurements with ss polarization, and the dashed lines represent the measurements with pp polarization.
The green line shows the specular reflectance defined by Eq.~(\ref{specular}), and the red line shows the total reflectance defined by Eq.~(\ref{diffuse}). 
The gray line shows the total reflectance of the direct coupling discussed in Sec.~\ref{directpath}. The vertical axis is different between the upper and lower panels.}
\label{AoI30_reflectance}
\end{figure*}
\par
Figure~\ref{AoI30_reflectance} and Table~\ref{tbl:reflectance} compare the specular reflectance of the absorbers $R^\text{s}$ in the S- and P-polarization, calculated using Eq.~(\ref{specular}). The specular reflectance of AN-72 is -14 dB for S-pol and -19 dB for P-pol at 90 GHz, which is 5 dB different for the two polarizations. 
On the other hand, the specular reflectance of HR-10 is -22 dB for the S-polarization, which is smaller than that of AN-72. 
In general, dielectric materials have more significant reflectance for the P-pol than the S-pol for oblique incident wave\cite{orfanidis2016}. 
The measurements of AN-72 show a consistent tendency.
By contrast, the reflectance of LS-22 shows little difference between the two polarizations. 
The reflectance of TK-L was about -40 dB for both polarizations, the lowest among the five absorbers measured in this experiment.
\par
Figure~\ref{AoI30_reflectance} and Table \ref{tbl:reflectance} also show the total reflectance $R^\text{t}$ calculated using Eq.~(\ref{diffuse}). Here, $\theta_\text{max}$ in Eq.~(\ref{diffuse}) is set to 30$^\circ$. While the specular reflectance (green line in Fig.~\ref{AoI30_reflectance}) of HR-10 for the S-polarization is -22 dB, it rises to -15 dB for total reflectance (red line). This shows that HR-10 has a larger diffuse reflection than other absorbers, consistent with the isotropic spreading feature of angular patterns in  Fig.~\ref{farfield_30}.
The total reflectance of TK-S is also higher than the specular reflectance by around 5 dB. For other absorbers, the two reflectances have little difference, indicating that diffuse reflection is small. 

\begin{table}
\caption{Specular and total reflectance of millimeter-wave absorbers at 90 GHz for AoI = 30$^\circ$.}
\begin{center}
\label{tbl:reflectance}
\begin{tabular}{lcccc}
\hline
& \multicolumn{2}{c}{ss} & \multicolumn{2}{c}{pp} \\
Samples & Specular & Total & Specular & Total \\
\hline
AN-72 & -14 dB& -14 dB & -19 dB & -18 dB\\
HR-10 & -22 dB & -15 dB & -21 dB & -19 dB\\
LS-22 & -11 dB & -11 dB & -12 dB & -12 dB\\
TK-L & -43 dB & -39 dB & -45 dB & -38 dB\\
TK-S & -25 dB & -21 dB & -25 dB & -20 dB\\
\hline
(direct coupling) & -65 dB & -54 dB & -56 dB & -41 dB\\
\hline
\end{tabular}
\end{center}
\end{table}

\subsection{Effects of direct coupling} 
\label{directpath}
In the conventional horn-to-horn methods, the minimum measurable reflectance is limited by the direct coupling from the transmitting horn to the receiving horn, especially for measurements with larger angles of reflection\cite{saily2003}. 
\par
We found that the near-field method has similar limitations. The measured angular responses for AoI = 45 $^{\circ}$ (Fig.~\ref{farfield_45}) show a similar feature with vertical stripes at $\theta\cos{\phi} \sim -20^\circ$, regardless of the test absorber. 
This is because of interference between the reflected waves and the direct coupling from the lens horn to the probe. Such a feature is more apparent in the P-pol measurements, which is consistent with the higher sidelobe level of the lens horn in its polarization direction.
\par
We evaluated the direct coupling in our experimental setup by performing near-field measurements without placing the millimeter-wave absorbers. The results are shown in Fig.~\ref{AoI30_reflectance} (gray lines) and in Table~\ref{tbl:reflectance}. In the case of P-pol measurements, the measured total reflectance of TK-L is close to the direct coupling, implying that the measured value may only represent an upper limit.
\par
In future experiments, the accuracy of the measurement may be improved by reducing the direct coupling. For example, a horn with a small sidelobe such as a corrugated horn could be used. Also, the time-domain filtering in the near-field measurement\cite{Takakura2022} can be helpful to distinguish the direct coupling from the reflected waves from the absorber and remove the effects of the direct coupling.

\subsection{Comparison with horn-to-horn method}
\begin{figure}
\centering
\includegraphics[width=\linewidth]{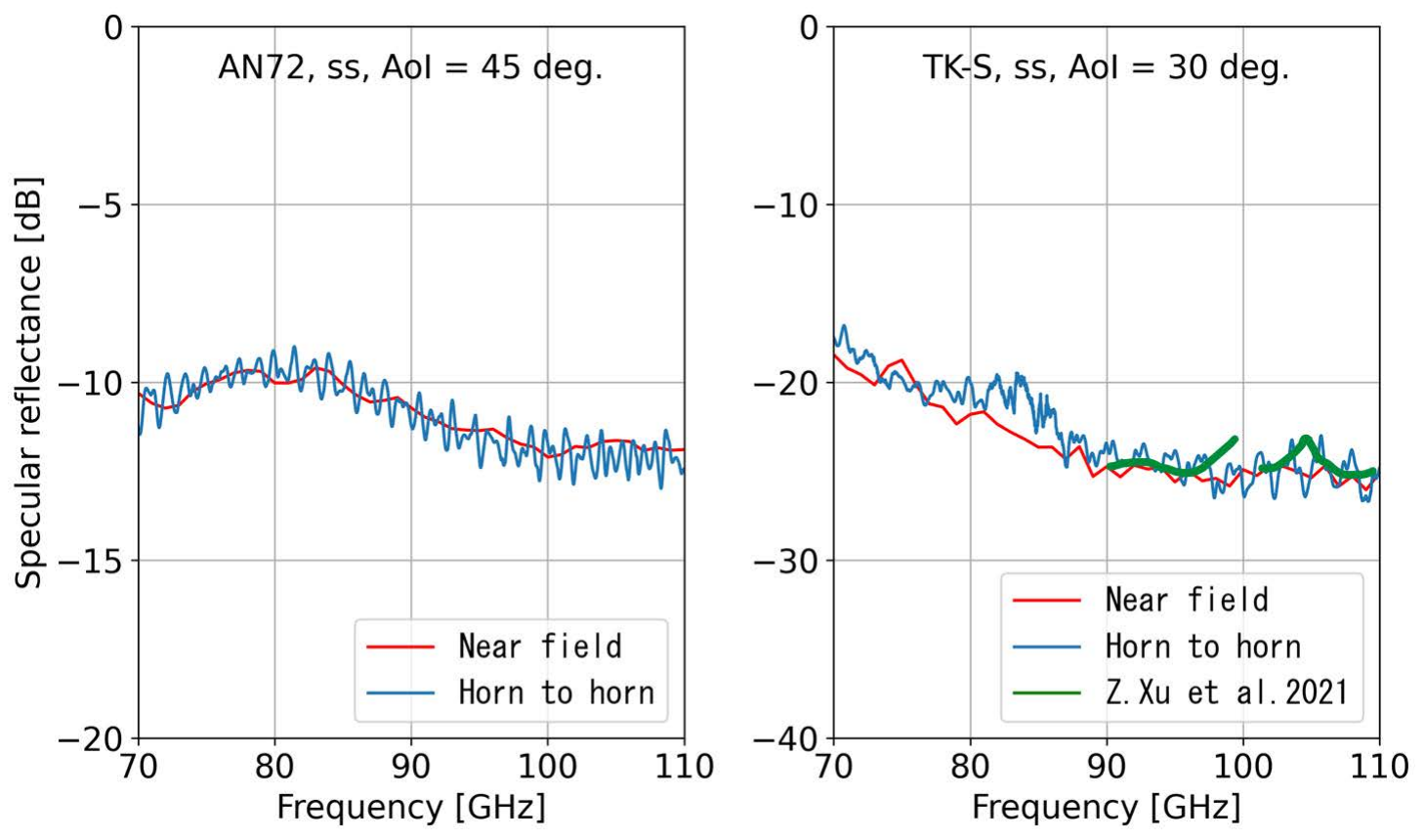}
\caption{Comparison of planar near-field reflection measurements and specular reflectance measured by the \textit{horn-to-horn} method. 
The left panel shows the reflectance of Eccosorb AN-72 with ss polarization for AoI = 45$^\circ$. 
The right panel shows the reflectance of TK-S with ss polarization for AoI = 30$^\circ$.
The red lines represent reflectance derived from Eq.~(\ref{specular}) using planar near-field measurements, and the blue lines represent reflectance measured by the horn-to-horn method. The blue line is normalized by the reference aluminum plate measured by the horn-to-horn method. For the reflectance of TK-S, the green line shows the horn-to-horn measurements of Z. Xu et al. 2021\cite{Xu:21} at 90 GHz -- 110 GHz.}
\label{AN72}
\end{figure}

We compared the measured specular reflectance discussed in Sec.~\ref{horntohorn} with conventional horn-to-horn measurements. 
We performed the horn-to-horn measurements using the same experimental setup described in Sec.~\ref{experimentalsetup} by replacing the probe with the lens horn. The two lens horns for the transmitter and receiver are identical.
The results demonstrate that the planar near-field measurements and the horn-to-horn reflection measurements agree within a few dB and that the former are less affected by standing waves (Fig.~\ref{AN72}). The measurements are consistent with the specular reflectance of TK-S reported in Ref.~\cite{Xu:21}, measured with the horn-to-horn method at 90 -- 110 GHz. 
\par
On the other hand, compared to the horn-to-horn method, the near-field method takes longer for each measurement. Also, near-field measurements require a wide scan plane and thus need a larger mechanical clearance between the transmitter and receiver, which limits the measurable angle of incidence.

\subsection{Electric field measurements perpendicular to the co-polarization of the lens horn}
\begin{figure*}
\centering\includegraphics[width=\linewidth]{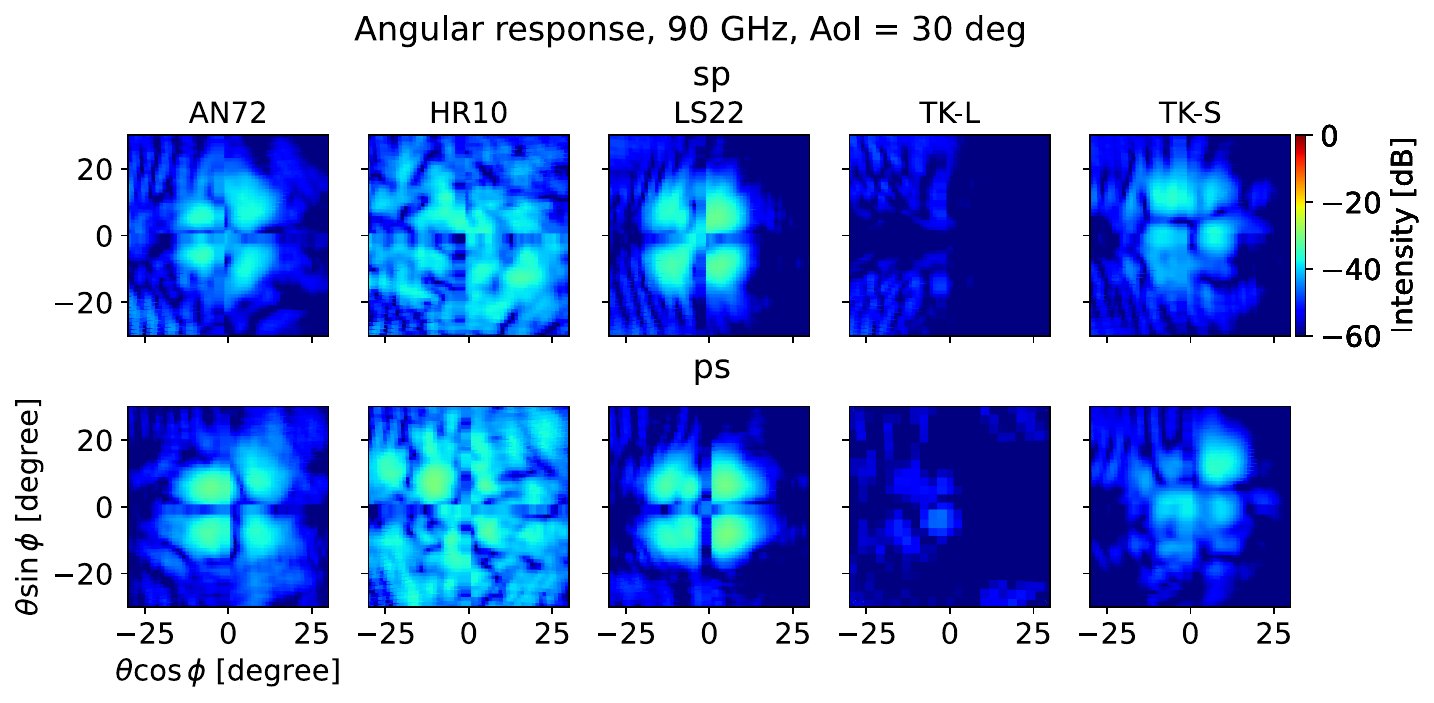}
\caption{Two-dimensional reflection pattern of the electric field perpendicular to co-polarization of the lens horn at 90 GHz for AoI = 30$^{\circ}$.
The label ``sp'' represents the measurements with the S-polarized lens horn and the P-polarized probe horn, and vice versa for ``ps''.
The ``sp'' and ``ps'' patterns are normalized by the peak value of the ``ss'' and ``pp'' measurements of the reference aluminum plate, respectively.}
\label{farfield_crosspol}
\end{figure*}

\begin{figure*}
\centering\includegraphics[width=15cm]{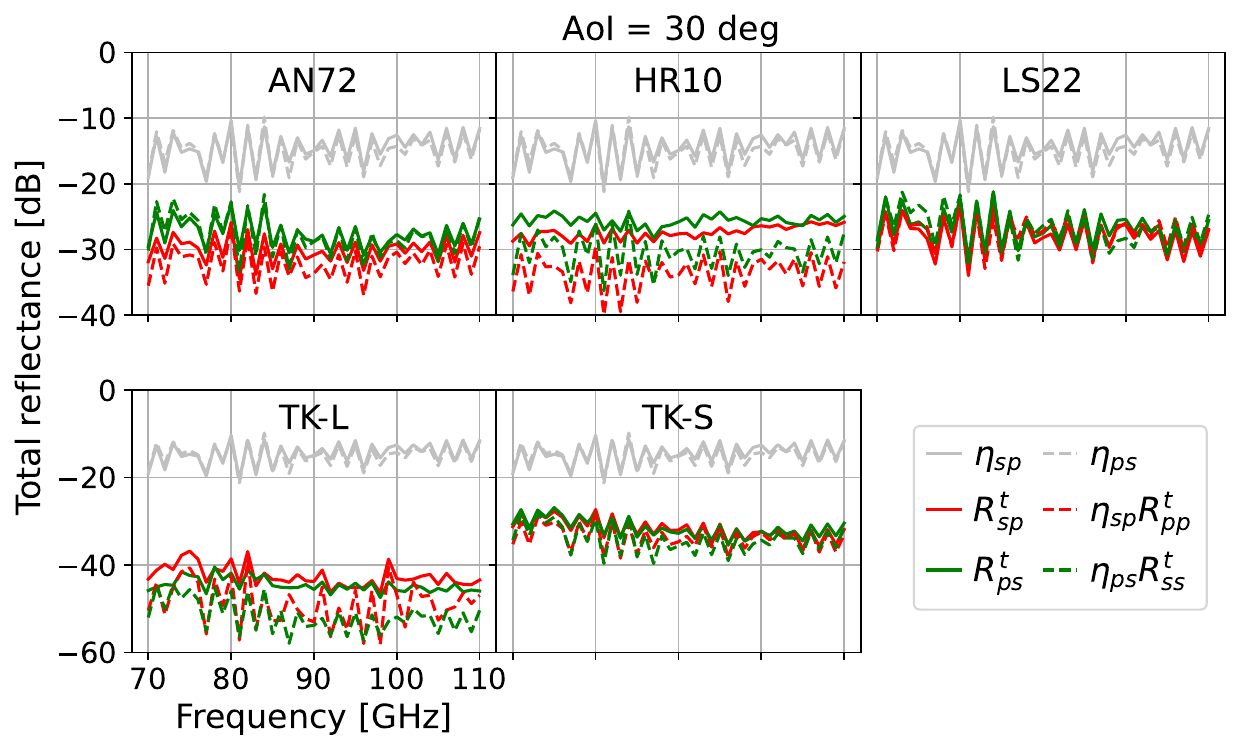}
\caption{Total reflectance of the electric field perpendicular to co-polarization of the lens horn for AoI = 30$^\circ$. The solid red line shows the reflectance with sp polarization, and the solid green line shows the reflectance with ps polarization. $R^{t}_{sp}$ and $R^{t}_{ps}$ shown by solid lines include $\eta_{sp} R^{t}_{pp}$ and $\eta_{ps} R^{t}_{ss}$ shown by dashed lines.}  
\label{AoI30_reflectance_cross}
\end{figure*}

Using the planar near-field measurement method, we also measured the electric field perpendicular to the co-polarization of the lens horn.
Figure~\ref{farfield_crosspol} shows the angular responses of the electric field perpendicular to the co-polarization of the lens horn at 90 GHz for AoI = 30$^{\circ}$.
The patterns of AN-72, LS-22, and TK-S measured with the S- and P-polarized lens horn (labeled as ``sp'' and ``ps'' respectively) are similar to those of the reference aluminum plate (``sp'' and ``ps'' in Fig.~\ref{cal_Al}) with attenuated levels, showing quadrupoles. This indicates that the patterns are determined mainly by the cross-polarization of the lens horn. On the other hand, the patterns of HR-10 and TK-L do not resemble that of the reference aluminum plate.
\par
Such trends are also apparent in total reflectance defined by Eq.~(\ref{diffuse}), as summarized in Fig.~\ref{AoI30_reflectance_cross}. 
For AN-72, LS-22, and TK-S, the total reflectance measured with the S- and P-polarized lens horn ($R_{sp}^{t}$ and $R_{ps}^{t}$ respectively) are primarily consistent with $\eta_{sp} R^{t}_{pp}$ and $\eta_{ps} R^{t}_{ss}$. Here, $\eta_{sp}$ and $\eta_{ps}$ indicate the cross-polarization components of the lens horn, which are obtained from the reference aluminum plate measurements as, 
\begin{equation}
    \eta_{sp} = \frac{\int^{\theta=30^\circ}_{\theta=0^\circ}I^{\text{Al}}_{sp}(\theta,\phi)d\Omega}{\int^{\theta=30^\circ}_{\theta=0^\circ}I^{\text{Al}}_{ss}(\theta,\phi)d\Omega}, \quad
    \eta_{ps} = \frac{\int^{\theta=30^\circ}_{\theta=0^\circ}I^{\text{Al}}_{ps}(\theta,\phi)d\Omega}{\int^{\theta=30^\circ}_{\theta=0^\circ}I^{\text{Al}}_{pp}(\theta,\phi)d\Omega}.
    \label{eta}
\end{equation}
In contrast, the total reflectance $R_{sp}$ and $R_{ps}$ of HR-10 and TK-L are $\sim 5$ dB larger than  $\eta_{sp} R_{pp}$ and $\eta_{ps} R_{ss}$. These measurements imply that HR-10 and TK-L may affect the polarization direction when reflecting the incident waves.

\section{Conclusion}
\label{conclusion}
By applying the planar near-field antenna measurement method, we have developed a method to measure the reflectance of millimeter-wave absorbers, including the two-dimensional distribution of diffuse reflections. 
We have measured five absorbers (TK-L, TK-S, and Eccosorb AN-72, HR-10, LS-22) at $30^\circ$ and $45^\circ$ angles of incidence for two polarizations (S-pol and P-pol) in the frequency range from 70 GHz to 110 GHz.
From these measurements, we evaluate the specular and total reflectance of each absorber. 
The obtained specular reflectance is less affected by standing waves and agrees with the conventional measurement. 
The specular reflectance of AN-72 and HR-10 are higher for S-pol, while that of LS-22, TK-L, and TK-S does not depend on incident polarization. 
We also evaluate the total reflectance, including diffuse reflection. 
The diffuse-to-specular reflectance ratio is significantly higher for HR-10 and TK-S compared with AN-72 and LS-22.
We also measured the electric field perpendicular to the co-polarization of the lens horn. The measurements imply that HR-10 and TK-L may affect the polarization direction when reflecting the incident waves.
\par
While the conventional horn-to-horn method measures the optical coupling efficiency between the two horns, the present method allows the detailed near-field mapping of the wave reflected from the absorber. 
Thus, the latter can investigate the nature of the reflection more extensively. 
We have demonstrated in this paper that the diffuse-to-specular reflectance ratio is different among various materials. 
It is also a merit of the new method that the measurement is not sensitive to the alignment of the absorber in the setup optics. 
Practically, it will help us conduct similar reflection measurements in cryogenic conditions. 
The method is expected to contribute to the development of millimeter-wave absorbers by achieving accurate measurements, including diffuse reflections, and to make a better design against the stray light in CMB observation telescopes.

\textbf{Funding}
This work was supported by MEXT/JSPS KAKENHI Grant Numbers 23K25889, JP23K17309, JP23K25889 and JP24K17078 and by JSPS Core-to-core Program Number JPJSCCA20200003.

\textbf{Acknowledgments}
We are grateful to Luca Lamagna, Andrea Occhiuzzi, Haruaki Hirose, Takuro Fujino, Rion Takahashi and Karen Sakamoto for their helpful discussions on the measurements. We are also grateful to Oshima Prototype Engineering Co. for fabricating the horn antennas used in the measurements.

\textbf{Disclosures}
The authors declare no conflicts of interest.

\textbf{Data Availability}
Data underlying the results presented in this paper are not publicly available at this time but may be obtained from the authors upon reasonable request.

\bibliography{ref.bib} 
\end{document}